# Social networks as an apparatus for managing of information security in the digital economy

A. A. Shyian

Vinnitsia National Technical University, Vinnitsia, Ukraine

**Abstract**
The paper aims to present a new apparatus for managing of the information security of the digital economy with using of social networks. A general problem for optimization of the information security management for participants in the digital economy is formulated. The results obtained make it possible to develop a new class of programs for analysis and decision support in the field of information security and economic management. The development of this line of research will be especially important in the financial field.

**Keywords:** digital economics; information security; social network; path integral; management.

Let's present the digital economy as such a mathematical model.

The "material (classical) part" of the digital economy consists of product manufacturers, logistics elements, warehouses (archives) and consumers. It can be represented by the following tuple.

$$MP = <M, C, L, D> \qquad (1)$$

The set of producers of a product will be denoted as a finite set $M$ (of cardinality $M$) in the set of natural numbers $N$. An individual producer will then be denoted as $m_i$, $i=1,\ldots,M$.

The set of consumers of the product will be denoted as a finite set $C$ (cardinality $C$) in the set of natural numbers $N$. A separate producer will then be denoted as $c_j$, $j = 1, \ldots, C$.

The set of logistics elements on the way from producer mi to consumer cj will be denoted as a finite set $L$ (of cardinality $L$) in the set of natural numbers $N$. An individual producer will then be denoted as $l_{p, i \to j}$, $i = 1, \ldots, M$, $j = 1, \ldots, C$, $p = 1, \ldots, L$.

The set of warehouses of the product will be denoted as a finite set $D$ (of cardinality $D$) in the set of natural numbers $N$. A separate warehouse will then be denoted as $d_{k, \{i\}, \{j\}}$, where $k = 1, \ldots, D$, and the set $\{i\}$ is a collection the producers that the warehouse works with, and the set $\{j\}$ is a collection of consumers who receive the goods from the warehouse.

This part of the economy has been fairly well researched and formalized [1, 2].

The informational part of the digital economy consists from producers, consumers, individuals and social groups that influence both the consumer and the producer. This influence is carried out through the channels of formation and transmission of information, in particular through schools, universities, culture, religion and the like. There are such instruments of influence: textbooks and monographs, fiction in various directions (from historical or religious texts to fairy tales, fantasy and science fiction), social networks, electronic content (texts, audio, video, photos, etc.), desktop, computer and network games, radio, television, theater, concerts and many others.

However, all these tools are based on a common source: it is the activity of the person who creates specific content. This content is influencing the other person. Such influence is carried out by changing the picture of the world for the individual.

Mathematically, this can be described as follows.

A specific person is modeled as a specific node in a social network. Communication between individual nodes is carried out using different tools (for example, those named above). Thus, we define a specific graph.

Let's set the nodes that correspond to producers and consumers.

Let us define a "path" between two nodes as a set of oriented links and nodes through which information is transmitted from one node (source) to another node (recipient). In general, there can be a lot of such paths. If we take into account the time factor, then there are a lot of such paths between two nodes. For example, it can be a waste of time not perceiving, analyzing and processing information ("thinking") information in intermediate nodes. Also, in a number of cases, texts with a multi-thousand-year history can act as a "source of information", the "waves" from which can capture hundreds of thousands (or even millions) of network nodes.

All the above set of nodes (that is, specific people who lived before and are living today) distorts the information that is transmitted to the network by the source. Other nodes "layer" their information on information from the source. Moreover, such layering often has specific goals.

As a result, we can, as a first approximation, represent the resulting influence of information on the recipient from the source in the form of a definite path integral [3, 4]. Integration is carried out along all paths along which information reaches the recipient. This should also take into account the arrays of information that are distributed over time.

$$I(S \to R, t) = F_0(S) + \int_{Source}^{Recipient} F[l(\tau)] Dl + F(R, t) \tag{2}$$

In (2), such notation was introduced. $I(S \to R, t)$ is the resultant influence of information from the source on the recipient as a result of information passing along all possible paths in the network by the time $t$. Function $F[l(\tau)]$ is a functional describing the change in information along one of the possible paths. $Dl$ is the standard differential in the path integral. Function $F_0(S)$ is information that the source launches into the network at $t=0$. Function $F(R, t)$ is the information that the recipient possesses at the time $t$.

It follows from (2) that the second term has a decisive influence on the change in the recipient's perception of economically important information. Thus, to optimize the external influence on the recipient of economic information, we come to such an optimization problem.

$$\arg\min_{F_1} \left\{ \int_{Source}^{Recipient} \{F[l(\tau)] - F_1[l(\tau)]\} Dl \right\} \tag{3}$$

Here the reaction of the function $F_1[l(\tau)]$ is calculated using the following formula.

$$F_1[l(\tau)] = \sum_i \int_{S_i}^{l(\tau)} F_c[l(\alpha)] Dl \qquad (4)$$

Here $S_i$ is a finite number of "agents of influence" on the nodes through which the reaction is carried out, $i=1,\ldots,S$. Function $F_c[l(\alpha)]$ is the influence of the counteraction, which the information security authorities exert on the network node, through which the path passes from the given source of the considered information to its given recipient.

Thus, we reduced the general task of implementing information security mechanisms in the digital economy (3) to finding the function $F_c[l(\alpha)]$. The general patterns that such functions should have are described below.

1. Social networks create opportunities for the formation of social groups that have an impact on both their members and the social community as a whole. Typically, such social groups are formed around a relatively small number of individuals. Such informal social formations, each of which consists of a relatively small number of people, can be called "areas of influence" in the social network. Thus, we get a certain area $A_q$ localized in the network, which is formed around a fairly small number of people (they can be called "centers of influence" or "leaders"). Each such area $a_q \in A_q$ can be characterized by those informational characteristics $CH(a_q)$ that the given area operates with.

2. The set of all paths from the source to the receiver can be divided into a finite set of paths $W$ that pass through each of the domains $a_q \in A_q$. It can be written this way.

$$W_h = \sum_{q=1}^{Q} W(a_q) + \sum_{q \neq b} W(a_q, a_b) + \ldots + W(a_1, a_2, \ldots, a_Q) \qquad (5)$$

In (5) it was recorded that some of the paths can pass through several areas of influence.

3. For specific tasks, certain characteristics are set that highlight only some of the areas of influence as those that may have an impact on a specific recipient. Therefore, in (5) only the "necessary" areas of influence should be taken into account.

4. A subset of the set $A_q^{inf} \subset A_q$ of areas of influence can be selected in the set $A_q$. They will be used in (4), and then in (4).

5. Finally, it is necessary to take into account the transit time of information from the source to the receiver, which is written explicitly in (3) and (4).

In conclusion, we note that for a more detailed management of the information security of the digital economy at the micro and macro levels, it is necessary to use a more detailed description of the set of paths $W$. In particular, it is necessary to use the apparatus of the theory of homotopy and homology [5 – 8], as well as multidimensional classifications for discrete networks [9]. This is necessary due to the fact that during the transition to macroeconomics, it becomes necessary to take into account the interaction of areas of influence with each other, as well as differences in education at different nodes of the social network. Moreover, some areas of influence include groups of scientists, teachers, and university students. Finally, an important circumstance is the presence of mutual influence between different areas of influence. For example, it can be religious social groups that have different history of communication (including even a categorical rejection of each other). Further research will allow developing a powerful conceptual apparatus and universal mathematical models that can be applied to a wide range of economic situations, primarily in the financial field. For example, it can be the optimization of financial mechanisms to accelerate the socio-economic development of Ukraine. Or the management of innovation processes, especially for their impact on the competitiveness of the national economy in the international arena. It is also interesting to use the results obtained to study the state and prospects of investment activity in modern Ukraine. And the problems of managing the potential of a modern enterprise, when the possibility for international collaboration especially would be taking into account, can be an interesting field of application of the developed conceptual and mathematical apparatus. The obtained results allow developing a new class of computer programs for analysis and decision support in the field of information security and economic management. The development of this line of research will be especially important in the financial field.

**Shyian Anatolii A.**, PhD in Physics and Mathematics, Associate Professor, Department of Management and Security of Information Systems, Vinnitsia National Technical University, Vinnitsia, E-mail: anatoliy.a.shiyan@gmail.com.